\documentclass[aps,prl,twocolumn,superscriptaddress,nofootinbib]{revtex4-2}
\usepackage[T1]{fontenc}
\usepackage{amsmath,amssymb,bm}
\usepackage{microtype}
\usepackage[dvipsnames]{xcolor}
\usepackage{tikz}
\definecolor{modgreen}{rgb}{0,0.45,0}
\definecolor{modred}{rgb}{0.45,0,0}
\newcommand{\new}[1]{#1}

\usepackage[colorlinks=true,citecolor=blue,urlcolor=blue]{hyperref}

\newcommand{\D}{\mathfrak D}
\newcommand{\dd}{\mathrm{d}}
\newcommand{\Favre}[1]{\widetilde{#1}}
\newcommand{\lFavre}[1]{\widehat{#1}}
\newcommand{\Srho}{{\cal I}^\rho}
\newcommand{\Adif}{\bm A}          
\newcommand{\Am}{\bm{\mathcal A}}  
\newcommand{\Cm}{\bm{\mathcal C}}  
\newcommand{\Lm}{\bm{\mathfrak L}} 
\newcommand{\bhat}{\hat{\bm b}}

\begin{document}
\title{Scalar Dissipation Criticality in Compressible Magnetized Turbulence}
\author{Svitlana Mayboroda}
\affiliation{Department of Mathematics, ETH Z\"urich, R\"amistrasse 101, 8092 Z\"urich, Switzerland}
\author{David N. Spergel}
\affiliation{Flatiron Institute, Simons Foundation, 162 Fifth Avenue, New York, New York 10010, USA}
\author{Camillo De Lellis}
\affiliation{School of Mathematics, IAS, Princeton, New Jersey 08540, USA, and GSSI, L'Aquila, Italy}
\date{August 23, 2026}

\begin{abstract}
We extend the Obukhov–Corrsin theory of scalar turbulence to compressible flows with spatially variable, anisotropic diffusivity. Combining density and diffusivity into a single positive matrix field, a transport landscape,  permits an exact scale-by-scale balance in which the sign-indefinite commutator between filtering and diffusion is eliminated rather than estimated. If the density-weighted third-order velocity and scalar increments scale as $\ell^\alpha$ and $\ell^\beta$, respectively, we prove that anomalous scalar dissipation is impossible when $\alpha+2\beta>1$. Remarkably, this threshold is independent of the anisotropy and spatial regularity of the diffusivity, provided it remains uniformly elliptic. Codimension-one shocks in both velocity and scalar have $\alpha=\beta=1/3$ and therefore lie exactly at the critical threshold. For statistically stationary turbulence we further obtain an exact density-weighted constant-flux relation, providing a compressible analogue of the relation underlying Yaglom’s law. The results apply directly to passive-scalar transport in both gases and magnetized plasmas and provide testable diagnostics for simulations of compressible magnetohydrodynamic turbulence.
Because the same
operator governs anisotropic heat conduction, the results carry over to the
electron temperature of a magnetized plasma, and they imply that gradient
statistics in such flows must be contracted with the transport landscape rather than with
the density.
\end{abstract}

\maketitle

Turbulent mixing of a tracer is a basic process in combustion
\cite{PoinsotVeynante2005} and atmospheric transport \cite{SeinfeldPandis2016}.
In plasmas the magnetic field makes the transport coefficient a \emph{tensor}.
Charged particles gyrate about field lines far faster than they collide, so
both heat and species transport occur preferentially along field lines, and
tracer transport in astrophysics is correspondingly anisotropic
\cite{deAvillezMacLow2002,PanScannapieco2010,ColbrookMaHopkinsSquire2017,PanScannapiecoScalo2013,Hopkins2017,Ruszkowski2011,SharmaCQP2009,HanaszLesch2003,Snodin2006,Pakmor2016},
as it is in fusion plasmas
\cite{Feng2004,Dai2016,Dux2006,Sciortino2021,Krasheninnikov2011,Pigarov2005}.  

The Obukhov--Corrsin--Yaglom theory \cite{Obukhov1949,Corrsin1951,Yaglom1949}
connects the transfer of scalar variance across scales to scalar dissipation in
incompressible turbulence, and its coarse-graining formulation is part of the
broader Onsager framework for anomalous dissipation
\cite{ConstantinTiti1994,DuchonRobert2000,Eyink2003}.  The theory predicts the
Yaglom law seen in simulations \cite{WatanabeGotoh2004} and experiment
\cite{Sreenivasan1996}.  Its compressible counterparts for the carrier flow are
the coarse-grained cascade and dissipation-anomaly results of
Refs.~\cite{Aluie2011,Aluie2013,GaltierBanerjee2011,BanerjeeGaltier2013,EyinkDrivas2018,DrivasEyink2018,FeireislGwiazda2017},
and the Lagrangian fluctuation--dissipation relation of
Ref.~\cite{DrivasEyink2017}.  Explicit constructions show that the
corresponding scalar thresholds can be crossed and that anomalous scalar
dissipation genuinely occurs above them
\cite{ColomboCrippaSorella2023,ArmstrongVicol2025,BrueDeLellis2023}.

This Letter extends these concepts to compressible transport with anisotropic diffusivity.  A geometric observation --- that the whole problem
lives on a Riemannian manifold built from the landscape --- motivates the
scale-by-scale balance \eqref{eq:master}, whose crucial terms are manifestly
positive.  Shock-dominated transport, specific to compressibility, turns out to
sit exactly on the critical line.  For gases, the landscape simplifies to a scalar field proportional to the product of the density and the viscosity and its treatment is a special case of the plasma analysis.\new{ We provide a more formal mathematical 
analysis of this problem in a companion paper\cite{Companion}}

\textit{Setting.}
On the periodic torus $\mathbb T^d$, $d\ge2$, normalized so that
$\int\rho\,\dd x=1$ for all $t$, we consider
\begin{align}
 \partial_t\rho+\nabla\!\cdot(\rho\bm v)&=0,
 \label{eq:cont}\\[-2pt]
\partial_t(\rho\theta)+\nabla\!\cdot(\rho\bm v\theta)
 &=\kappa\,\nabla\!\cdot\!\big(\rho\,\Adif\nabla\theta\big)+S .
 \label{eq:scalar}
\end{align}
\new{Here $\kappa$ is the diffusive amplitude sent to zero in the inviscid
limit}, $\Adif(x,t)$ is a dimensionless diffusivity profile, and $S(\theta,x,t)$
is a source.  
We assume of $\Adif$ only that it is measurable, symmetric, and
uniformly elliptic,
\begin{equation}
 a_-\bm I\preceq\Adif(x,t)\preceq a_+\bm I ,
 \qquad 0<a_-\le a_+<\infty ,
 \label{eq:ellipticity}
\end{equation}
in the sense of quadratic forms.  No continuity in $x$ or $t$, no bound on
$\nabla\Adif$, no isotropy, and no bound on the anisotropy ratio $a_+/a_-$ is
imposed, and no relation between $\Adif$ and $\rho$ is assumed.  The mass
normalization is what makes the coarse-grained density strictly positive
without any lower bound on $\rho$.

For a plasma with $\bhat=\bm B/|\bm B|$, the symmetric part of the Braginskii
closure \cite{Braginskii1965,SpitzerHarm1953,EpperleinHaines1986} has the
eigenframe of $\bhat$, and normalizing the amplitude by the maximum of the  parallel
coefficient, $\kappa=\chi_\parallel^{max}$, gives
\begin{equation}
 \Adif=\bhat\bhat+(\Omega\tau)^{-2}\big(\bm I-\bhat\bhat\big),
 \quad a_+=1,\;\; a_-\simeq (\Omega\tau)^{-2},
 \label{eq:braginskii}
\end{equation}
where $\chi_\parallel/\chi_\perp=(\Omega\tau)^2$ reaches $\sim10^{10}$ in
tokamak plasmas \cite{dCNChacon2011,Gunter2005,Holzl2009} and is larger still
in the intracluster medium.  This anisotropy is not a perturbation: it changes
the qualitative behavior of the medium, generating buoyancy instabilities
absent in the isotropic case
\cite{Balbus2000,Quataert2008,ParrishStone2005}, and  is notoriously delicate
numerically
\cite{SharmaHammett2007,SharmaHammett2011,Gunter2005,ChaconDCNHauck2014}.
\new{The antisymmetric Righi--Leduc/Hall term in the conductivity, $b_\wedge \times$,
describes transport perpendicular to the field lines, rather than  diffusion, and can be absorbed into the velocity term
without changing the analysis of this paper\cite{Companion}.}

\textit{The transport landscape and its metric.}
Density and diffusivity enter every diffusive quantity below only through the
single symmetric positive-definite matrix field
\begin{equation}
 \Lm(x,t):=\rho(x,t)\,\Adif(x,t),
 \label{eq:landscape}
\end{equation}
which we call the \emph{transport landscape}. 
In $d=3$ , this landscape generates a Riemannian metric,
\begin{equation}
g_{ij}=\big(\det\Lm\big)\big(\Lm^{-1}\big)_{ij},
 \qquad \sqrt{|g|}=\det\Lm ,
 \label{eq:metric_landscape}
\end{equation}
the unique choice of the form $g^{ij}\propto\Lm^{ij}$ for which the
divergence-form diffusion operator is the Laplace--Beltrami operator of $g$ up
to a strictly positive factor.  Inserting \eqref{eq:braginskii} gives the two
eigenvalues differ by $(\Omega\tau)^{2}$.
Displacement along $\bhat$ costs a factor
$(\Omega\tau)^{-1}$ less metric distance than displacement across it: the
landscape manifold is compressed along the field by exactly the Hall parameter,
so two points joined by a field line are metrically close however far apart
they are in space.  This is the precise geometric content of the statement that
field lines are the highways of a magnetized plasma.  The compressive part of
the advection and the fluctuation and anisotropy of $\Adif$ are absorbed into
the metric; the remainder is divergence-free, a decomposition that is not the
Helmholtz one.  The picture is analogous to the localization landscape theory
for the Schr\"odinger operator \cite{MFPNAS} and it is what motivated the
identity below, although, as the End Matter shows, the proof itself needs only
the symmetry and pointwise positivity of $\Lm$.

\textit{Exact energy balance and the scale-by-scale identity.}
Multiplying \eqref{eq:scalar} by $\theta$ and using \eqref{eq:cont} gives
\begin{equation}
\boxed{\;
 \frac12\frac{\dd}{\dd t}\int\rho\theta^2\,\dd x
 =-\kappa\!\int\!\nabla\theta\cdot\Lm\nabla\theta\,\dd x+\int\theta S\,\dd x ,
 \;}
 \label{eq:unfilteredenergy}
\end{equation}
the compressible, landscape-weighted analogue of the usual scalar-energy
balance.  '

We coarse-grain with a fixed admissible kernel family: $K_\ell$ is
smooth, strictly positive, of unit mass, and satisfies the pointwise
logarithmic-gradient bound $|\nabla K_\ell|\le C_K\ell^{-1}K_\ell$ together with
$\int K_\ell(y)|y|^q\,\dd y\le C_K\ell^q$ for $0<q\le3$.  We choose to use the periodization of $\exp[-\sqrt{1+|y|^2/\ell^2}\,]$. \new{Note that the two popular filters,  $|\nabla K|/K=|y|/\ell^2$ and a Gaussian filter
do not satisfy the constraints and should not be used for numerical evaluation of these equation.}
Write $f_\ell=K_\ell*f$,
\begin{equation}
 \Favre f_\ell=\frac{(\rho f)_\ell}{\rho_\ell},
 \qquad
 \tau^\rho_\ell(f,g)=(\rho fg)_\ell-\rho_\ell\Favre f_\ell\Favre g_\ell ,
 \label{eq:favre}
\end{equation}
and put $h_\ell=\Favre\theta_\ell$, $\bm V_\ell=\Favre{\bm v}_\ell$,
$S_\ell=K_\ell*S$.  

\new{
For $0\le t<T$ set\footnote{Our $\Pi^\rho_\ell$ carries the opposite sign
convention to the companion paper, with a compensating sign in $E$; the
resulting identities coincide.}
\begin{align}
D&:=\kappa\int_t^T\!\!\int\nabla\theta\cdot\Lm\nabla\theta\,\dd x\,\dd s ,\\
 \Pi^\rho_\ell(s)&:=-\!\int\tau^\rho_\ell(\bm v,\theta)\cdot\nabla h_\ell\,\dd x ,
 \label{eq:Pi}\\
 B^\rho_\ell(s)&:=\tfrac12\!\int\tau^\rho_\ell(\theta,\theta)\,\dd x ,
 \label{eq:B}\\
 \Delta Q_\ell&:=\int_t^T\!\!\int(\theta S-h_\ell S_\ell)\,\dd x\,\dd s  \\
 E:&=B^\rho_\ell(t)-B^\rho_\ell(T)+\int_t^T\Pi^\rho_\ell\,\dd s+\Delta Q_\ell,\textit{}
 \label{eq:source-defect}
\end{align}
Here $D$ is the landscape-weighted dissipation, $\Pi^\rho_\ell$ is the resolved scalar flux, $B^\rho_\ell$  is the Favre
variance defect, and $\Delta Q_\ell$ the unresolved source work.  There are three positive dissipation terms:  $H$ the resolved
gradient dissipation, $N$ the commutator dissipation, and $R$ the unresolved
landscape gradient variance:  
\begin{align}
 H&:=\kappa\!\int_t^T\!\!\int\nabla h_\ell\cdot\Lm_\ell\nabla h_\ell\,\dd x\,\dd s ,
 \label{eq:H}\\
 N&:=\kappa\!\int_t^T\!\!\int\Cm_\ell\cdot\Lm_\ell^{-1}\Cm_\ell\,\dd x\,\dd s ,
 \label{eq:N}\\
R&:=\kappa\!\int_t^T\!\!\int\tau^{\Lm}_\ell(\nabla\theta,\nabla\theta)\,\dd x\,\dd s .
 \label{eq:R}
\end{align}
Each is nonnegative because $\Lm_\ell$ and $\Lm_\ell^{-1}$ are positive
definite and by \eqref{eq:matrix-CS} in the End Matter; $N$ is the squared length of the
commutator in the \emph{dual} metric.
}

 The  master balance equation,
\begin{equation}
 \boxed{\;D+R+N=2E+H\;}
 \label{eq:master}
\end{equation}
is derived in the End Matter and the source-free case is developed in
full in the companion mathematical paper 
\cite{Companion}. Importantly, the sign-indefinite  contribution from the commutator $\Cm_\ell:=(\Lm\nabla\theta)_\ell-\Lm_\ell\nabla h_\ell$ is never estimated:
it is eliminated algebraically against the two positive quantities $R$ and $N$.
Thus,
\begin{equation}
 D\le2E+H .
 \label{eq:masterineq}
\end{equation}

\textit{Weighted Obukhov--Corrsin criterion.}
Define the weighted Favre oscillation, with
$\dd\nu_{\ell,x}(y)=K_\ell(y)\rho(x-y)\rho_\ell(x)^{-1}\dd y$,
\begin{equation}
 \Srho_{p,\ell}(f)^p
 =\int\rho_\ell(x)\!\int
 |f(x-y)-\Favre f_\ell(x)|^p\,\dd\nu_{\ell,x}(y)\,\dd x ,
 \label{eq:S}
\end{equation}
a $p$-th oscillation with respect to a probability measure on the pair $(x,y)$.
For a family $(\rho,\bm v^{(\kappa)},\theta^{(\kappa)})$ assume,
uniformly in $\kappa$ and for some $\alpha,\beta\in(0,1]$ and all
$\ell\le\ell_0$,
\begin{equation}
 \Srho_{3,\ell}(\bm v^{(\kappa)})\le M_v\ell^\alpha,
 \qquad
 \Srho_{3,\ell}(\theta^{(\kappa)})\le M_\theta\ell^\beta .
 \label{eq:scaling}
\end{equation}
No hypothesis of any kind is placed on  $\Lm$: the diffusivity may
be rough, discontinuous, or fluctuate on the smallest scales, and $\bhat$ may
be turbulent or chaotic.  
For a nonzero source write $S=\rho f$
and assume in addition
$\Srho_{2,\ell}(f)\le M_f\ell^{\gamma_f}$ with $\gamma_f>0$;
the source defect is then not an independent assumption, since the End Matter
gives $|\Delta Q_\ell(0,T)|\lesssim TM_fM_\theta\ell^{\gamma_f+\beta}$.

The coarse-graining estimates derived in the End Matter are
\begin{align}
 |\Pi^\rho_\ell|
 &\le C_K\ell^{-1}\Srho_{3,\ell}(\bm v)\big[\Srho_{3,\ell}(\theta)\big]^2
 \le C_KM_vM_\theta^2\ell^{\alpha+2\beta-1},
 \label{eq:fluxlemma}\\
 B^\rho_\ell&\le\tfrac12 M_\theta^2\ell^{2\beta},
 \qquad
 H\le C_K^2a_+T\,M_\theta^2\,\kappa\,\ell^{2\beta-2}.
 \label{eq:BH}
\end{align}
The exponent $\alpha+2\beta-1$ arises from exactly one velocity increment, two
scalar increments, and one derivative at scale $\ell$.  Neither the diffusivity
nor the landscape appears in \eqref{eq:fluxlemma} or in the bound on
$B^\rho_\ell$; the bound on $H$ is the only step of the whole argument at which
an upper bound on the diffusivity is used quantitatively.  We balance the bounds on the  flux
and the resolved gradient term by choosing
\begin{equation}
 \ell_\kappa=\kappa^{1/(1+\alpha)},
 \qquad \kappa\le\kappa_0:=\min\{1,\ell_0^{1+\alpha}\},
 \label{eq:ellk}
\end{equation}
Recall $\kappa =\chi_\parallel$. Using $\alpha\le1$ to make the endpoint contribution
$\ell_\kappa^{2\beta}$ subdominant, \eqref{eq:masterineq} gives
\begin{equation}
\begin{aligned}
 \kappa\!\int_0^T\!\!\int\nabla\theta^{(\kappa)}\!\cdot\Lm^{(\kappa)}
 \nabla\theta^{(\kappa)}\,&\dd x\,\dd t
 \le C\Big[\kappa^{\frac{\alpha+2\beta-1}{1+\alpha}}\\
 &+TM_fM_\theta\,\kappa^{\frac{\gamma_f+\beta}{1+\alpha}}\Big],
\end{aligned}
\label{eq:main}
\end{equation}
with $C$ affine in $a_+$, independent of $a_-$, and independent of $\kappa$;
for $S=0$ the second term is absent.  Since $\Lm\succeq a_-\rho\bm I$
pointwise, dividing by $a_-$ converts the left-hand side into the plain
density-weighted dissipation, so that
\begin{equation}
 \alpha+2\beta>1
 \;\Longrightarrow\;
 \lim_{\kappa\downarrow0}\kappa\!\int_0^T\!\!\int
 \rho|\nabla\theta^{(\kappa)}|^2\,\dd x\,\dd t=0 .
 \label{eq:limit}
\end{equation}
Equivalently, anomalous dissipation in either sense forces
$\alpha+2\beta\le1$.  The conclusion is 
one-sided: it neither asserts a Yaglom $4/3$ law nor sharpness at
$\alpha+2\beta=1$, since the signed mixed flux may be smaller than the product
of absolute oscillations through decorrelation, misalignment with $\nabla
h_\ell$, or spatial sign cancellation. We expect the rate in \eqref{eq:main} to be sharp nonetheless.

\new{
The analogue of \eqref{eq:limit} is well-known in the incompressible theory.
\cite{DrivasElgindiIyerJeong2022}.  By Favre-weighting the scalar and velocity fields (see \eqref{eq:scaling}, we can overcome the effects of compressibility and  by landscape weighting the  effects of spatially variable anisotropic dissipation  and extend this result to the more general case.}

\textit{Criticality of shocks.}
The increment hypothesis does not exclude shocks, and it places them exactly on
the critical line.  For a codimension-one discontinuity with an $O(1)$ jump,
only an $O(\ell)$ fraction of the mass contributes $O(1)$ increments at
separation $\ell$, so $\int|\delta_\ell\bm v|^p\,\dd x\sim\ell$ and
$\|\delta_\ell\bm v\|_{L^p}\sim\ell^{1/p}$.  What is needed of the density is
not a global bound but a non-degeneracy condition at the jump set: the mass
distribution pushed forward onto the jump normal must neither concentrate nor
vanish at the natural rate, $c_*r\le m((p-r,p))$,
$c_*r \le m(p,p+r)$, $m([p-r,p])\le C_*r$, and
$m([p,p+r] \le C_* r$.
Under that condition the weighted third-order structure function of such a
field is bounded above \emph{and below} by $r^{1/3}$, so the field lies in the
weighted Besov--Nikolskii class of order $\tfrac13$ and in no better one
\cite{Companion}.  Hence a generic shock satisfies \eqref{eq:scaling} with
$\alpha=\tfrac13$ and with no larger exponent, and the same counting applied to
a scalar with an $O(1)$ codimension-one jump gives $\beta=\tfrac13$, whence
\begin{equation}
 \alpha+2\beta=\tfrac13+\tfrac23=1 .
 \label{eq:shock-critical}
\end{equation}
A jump in the field direction, and hence in the eigenframe of $\Adif$,
coincident with the velocity shock --- the generic magnetized case --- affects
prefactors only, and the diffusivity is allowed to be discontinuous across the
shock, as it is whenever it depends on temperature.  Since $\alpha$ and $\beta$
are defined purely through velocity and scalar increments, shock criticality
survives an elliptic matrix $\Adif(x,t)$ of 
arbitrarily roughness.

The pair $\alpha=\beta=\tfrac13$ is also the classical
Kolmogorov--Obukhov--Corrsin pair of smooth incompressible turbulence.
Compressibility might have been expected to degrade it, since shocks are the
strongest singularities a compressible velocity field can support and a coarser
velocity admits a larger scalar flux; it does not.  The threshold is reproduced
by a completely different mechanism, and at the borderline our estimate neither
rules out nor establishes anomalous dissipation.  The exponent $3$ in
\eqref{eq:S} is precisely the one for which shocks are borderline, which is a
further reason for that choice.

\textit{Numerical target.}
The hypotheses \eqref{eq:scaling} refer to the coarse-graining kernel, but they
follow from membership of $\bm v$ and $\theta$ in a weighted
Besov--Nikolskii space defined with no kernel and no scale, through the
third-order structure function
$D^\rho_3(f;h)=(\int\rho|f(\cdot+h)-f|^3)^{1/3}$ measured in the mass
distribution \cite{Companion}.  That is the density-weighted structure function
already reported in the compressible turbulence literature, and it is what
makes a direct test possible.  Kritsuk \emph{et al.} found a third-order
exponent close to unity for $\rho^{1/3}\bm v$ at Mach 6, a cube-root exponent
near $\tfrac13$ \cite{Kritsuk2007}; their statistic is related to but not
identical with $\Srho_{3,\ell}(\bm v)$, and their mass dimension
$D_m\simeq2.4$ cautions that the dissipative geometry is more complicated than
a single codimension-one shock set.  Existing studies of compressible scalar
mixing show substantial Mach- and Schmidt-number dependence
\cite{Ni2015,Ni2016,PanScannapieco2010}.  Measuring $\Srho_{3,\ell}(\bm v)$ and
$\Srho_{3,\ell}(\theta)$ simultaneously would test whether shock-dominated
transport approaches $(\alpha,\beta)=(\tfrac13,\tfrac13)$.  \new{ Because the
criterion is insensitive to the diffusivity profile, such a test is not
confined to constant-Schmidt-number codes, nor to hydrodynamic ones: it applies
unchanged to magnetohydrodynamic simulations\cite{Hopkins2017,Ruszkowski2011}.
The landscape-weighted
dissipation is the physically meaningful quantity, and the unweighted, usually reported in simulations,
$\chi_{\parallel}\!\int\!\rho|\nabla\theta|^2$ overcounts by the anisotropy because it
charges perpendicular gradients at the parallel rate. }


In the strong-anisotropy limit, routinely
adopted in fusion modeling, $\chi_\perp/\chi_\parallel\to0$  and heat is
transported only along field lines
\cite{dCNChacon2011,dCNChacon2012,ChaconDCNHauck2014} As $a_-\to0$ at
fixed $a_+$, so $\det\Lm\to0$ and $g_\parallel\to0$: the metric distance along a
field line collapses and an entire field line degenerates to a point of the
landscape manifold.  At $a_-=0$ the filtered landscape may fail to be
invertible, so $\Am_\ell$, $\Cm_\ell$ and $N$ lose their meaning and the
commutator elimination breaks down; what survives is only the inequality
$G\le D$ in a convex-conjugate form \cite{Companion}, and with it neither
\eqref{eq:main} nor \eqref{eq:limit}.  This is a faithful rendering of the
physics rather than a defect: when field lines are chaotic and each collapses
to a point, the quotient of space by field lines is not a three-dimensional
manifold, and there is no reason to expect effective perpendicular transport to
be diffusive at all.  That is what is found numerically, where radial transport
in a fully chaotic field is reported to be non-diffusive with multivalued
flux--gradient relations \cite{dCNChacon2012}, against the quasilinear
expectation of Ref.~\cite{RechesterRosenbluth1978}.  The identity locates the
failure of the diffusive closure at the loss of rank of the landscape, not at
any property of the velocity field.

\textit{Stationary constant-flux relation.}
Assuming stationarity, \eqref{eq:unfilteredenergy} gives
\begin{equation}
 \D_\kappa^\rho
 :=\kappa\,\mathbb E\!\int\nabla\theta\cdot\Lm\nabla\theta\,\dd x
 =\mathbb E\!\int S\theta\,\dd x ,
 \label{eq:stationary-dissipation}
\end{equation}
so the mean landscape-weighted dissipation rate equals the mean injection rate.
Writing $J_\ell:=\kappa\!\int\nabla h_\ell\cdot\Lm_\ell\Am_\ell\,\dd x$ and
using the resolved budget of the End Matter together with
$\tau^\rho_\ell(f,\theta)=(S\theta)_\ell-h_\ell S_\ell$ yields the exact
stationary scale-by-scale identity
\begin{equation}
 \boxed{\;
 \mathbb E\Pi_\ell^\rho
 =\D_\kappa^\rho
 -\mathbb E\!\int\tau_\ell^\rho(f,\theta)\,\dd x
 -\mathbb EJ_\ell \;}
 \label{eq:stationary-master}
\end{equation}
Assume a nonzero
landscape-weighted anomaly, $\D_\kappa^\rho\to\D^\rho\in(0,\infty)$, and let
$r_\kappa\downarrow0$ as $\kappa\downarrow0$.  The forcing correction vanishes
under the source hypothesis, and the End Matter shows that $\mathbb
EJ_{r_\kappa}\to0$ provided $a_+\kappa\,r_\kappa^{2\beta-2}\to0$. Then
\begin{equation}
 \boxed{\;\mathbb E\Pi_{r_\kappa}^{\rho^{(\kappa)}}\longrightarrow\D^\rho .\;}
 \label{eq:constant-flux}
\end{equation}
In equilibrium, the mean density-weighted scalar-variance flux, $\Pi^\rho_l$, at scales separated from both the
forcing and the diffusive scale, equals to the common injection and dissipation
rate.  In the incompressible fluid limit,
$\Pi^\rho_l$ reduces. to $l^{-1} \delta_l v (\delta_l \theta)^2$, the familar incompressible scalar-variance flux.

\textit{Anisotropic heat conduction.}
The electron energy equation with Braginskii conduction $\bm\chi$,
\begin{equation}
 \tfrac32\partial_t(nT)+\nabla\!\cdot\!\big(\tfrac32nT\bm v\big)
 +nT\,\nabla\!\cdot\bm v
 =\nabla\!\cdot\!\big(n\,\bm\chi_T\nabla T\big)+Q ,
 \label{eq:heat-dictionary}
\end{equation}
with $Q$ a heating or radiative-loss rate, can be recast as \eqref{eq:scalar}
\emph{exactly} on multiplying by $\tfrac23$ by setting
$\rho:=n$, $\theta=T$, $\kappa\Adif:=\tfrac23\bm\chi_T$, and 
 $S:=\tfrac23\big(Q-nT\,\nabla\!\cdot\bm v\big).$
The compressional term is of the admissible form $S=\rho f$ with
$f=\tfrac23(Q/n-T\nabla\!\cdot\bm v)$.  Consequently
\eqref{eq:unfilteredenergy}, \eqref{eq:master}, \eqref{eq:fluxlemma} and
\eqref{eq:stationary-master} all hold for the electron temperature of a
magnetized plasma, with the landscape 
incorporating the physics of anisotropic heat conduction.

\new{There are two caveats to applying the scalar results to temperature:  First, the
source-increment hypothesis now contains$\nabla\!\cdot\bm v$.  At a
shock,  we can't assume  $\gamma_f>0$ as there is compressive heating at the shock,  so 
\eqref{eq:shock-critical} does \emph{not} apply to the temperature field.
Where compressional heating is weak relative to conduction, as in
the conduction-dominated core of the intracluster medium, the caveat is vacuous.
Second, the behavior of  $\Srho_{3,\ell}(\theta)$, one of the inputs to our derivation,
has to be tested for temperature in a particular set-up.}

\textit{Conclusions.}
We have given exact identities and an analogue of the
Obukhov--Corrsin--Yaglom theory for a compressible scalar with an arbitrary
symmetric positive-definite matrix diffusivity, so that the theory covers
anisotropic transport in magnetized media.  The deterministic criterion
\eqref{eq:limit} rules out anomalous weighted dissipation when
$\alpha+2\beta>1$, at a rate that is attained; stationarity yields the
compressible constant-flux result \eqref{eq:constant-flux}; and shocks lie
exactly on the critical line.  The landscape $\Lm=\rho\Adif$ generates
the metric \eqref{eq:metric_landscape}, whose anisotropy is the Hall parameter; it is the
weight under which the diffusive flux filters exactly; and it supplies the
inner product in which every dissipation-like term of \eqref{eq:master} is a
length.

Numerical studies of compressible turbulence
\cite{Ni2015,Ni2016,ColbrookMaHopkinsSquire2017} have struggled to find scaling
laws for scalar mixing.  Our work shows  gradient statistics should be weighted by
the landscape rather than by the density --- a distinction that matters for
codes with temperature-dependent or sub-grid transport coefficient..  In a magnetized
run the distinction is larger still: gradient statistics must be contracted
with $\Lm$, which weights field-parallel and field-perpendicular gradients by
coefficients differing by many orders of magnitude, and reporting
$\chi_\parallel\!\int\!\rho|\nabla\theta|^2$ instead overstates the dissipation
by the anisotropy ratio.  Future simulations should measure
$\Srho_{3,\ell}(\bm v)$, $\Srho_{3,\ell}(\theta)$ and $\Pi^\rho_\ell$, the
appropriate invariant quantities for arbitrary compressible density and matrix
diffusivity.

\begin{acknowledgments}
We acknowledge the Simons Foundation for support (BD Targeted-00017375).  DNS
thanks the Institute for Theoretical Studies, ETH, for its hospitality, and we
all thank CIEM (International Centre for Mathematical Meetings) in Castro
Urdiales for hospitality.  \textit{Data availability.}  No new data were
created or analyzed in this theoretical study.
\end{acknowledgments}
\bibliographystyle{apsrev4-2}

\bibliography{References}

@book{PoinsotVeynante2005,
  author    = {Poinsot, Thierry and Veynante, Denis},
  title     = {Theoretical and Numerical Combustion},
  edition   = {2},
  publisher = {R. T. Edwards},
  address   = {Philadelphia},
  year      = {2005}
}

@book{SeinfeldPandis2016,
  author    = {Seinfeld, John H. and Pandis, Spyros N.},
  title     = {Atmospheric Chemistry and Physics: From Air Pollution to Climate Change},
  edition   = {3},
  publisher = {Wiley},
  address   = {Hoboken},
  year      = {2016}
}

@article{deAvillezMacLow2002,
  author  = {de Avillez, Miguel A. and Mac Low, Mordecai-M.},
  journal = {Astrophysical Journal},
  volume  = {581},
  pages   = {1047},
  year    = {2002}
}

@article{PanScannapieco2010,
  author  = {Pan, L. and Scannapieco, Evan},
  journal = {Astrophysical Journal},
  volume  = {721},
  pages   = {1765},
  year    = {2010}
}

@article{ColbrookMaHopkinsSquire2017,
  author  = {Colbrook, M. J. and Ma, X. and Hopkins, Philip F. and Squire, Jonathan},
  journal = {Monthly Notices of the Royal Astronomical Society},
  volume  = {467},
  pages   = {2421},
  year    = {2017}
}

@article{PanScannapiecoScalo2013,
  author  = {Pan, L. and Scannapieco, E. and Scalo, J.},
  journal = {Astrophysical Journal},
  volume  = {775},
  pages   = {111},
  year    = {2013}
}

@article{Hopkins2017,
  author  = {Hopkins, Philip F.},
  journal = {Monthly Notices of the Royal Astronomical Society},
  volume  = {466},
  pages   = {3387},
  year    = {2017}
}

@article{Ruszkowski2011,
  author  = {Ruszkowski, Mateusz and Lee, Dongwook and Brueggen, Matthias and Parrish, Ian and Oh, S. Peng},
  journal = {Astrophysical Journal},
  volume  = {740},
  pages   = {81},
  year    = {2011}
}

@article{SharmaCQP2009,
  author  = {Sharma, Prateek and Chandran, Benjamin D. G. and Quataert, Eliot and Parrish, Ian J.},
  journal = {Astrophysical Journal},
  volume  = {699},
  pages   = {348},
  year    = {2009}
}

@article{HanaszLesch2003,
  author  = {Hanasz, M. and Lesch, H.},
  journal = {Astronomy and Astrophysics},
  volume  = {412},
  pages   = {331},
  year    = {2003}
}

@article{Snodin2006,
  author  = {Snodin, A. P. and Brandenburg, Axel and Mee, A. J. and Shukurov, Anvar},
  journal = {Monthly Notices of the Royal Astronomical Society},
  volume  = {373},
  pages   = {643},
  year    = {2006}
}

@article{Pakmor2016,
  author  = {Pakmor, R. and Pfrommer, C. and Simpson, C. M. and Kannan, R. and Springel, Volker},
  journal = {Monthly Notices of the Royal Astronomical Society},
  volume  = {462},
  pages   = {2603},
  year    = {2016}
}

@article{Feng2004,
  author  = {Feng, Y. and Sardei, F. and Grigull, P. and McCormick, K. and Kisslinger, J. and Reiter, Detlev},
  journal = {Plasma Physics and Controlled Fusion},
  volume  = {53},
  number  = {2},
  pages   = {024009},
  year    = {2011},
  note    = {The citation key is retained from the manuscript; publication year/volume correspond to the supplied bibliographic data}
}

@article{Dai2016,
  author  = {Dai, S. Y. and others},
  journal = {Contributions to Plasma Physics},
  volume  = {56},
  pages   = {130},
  year    = {2016}
}

@techreport{Dux2006,
  author      = {Dux, R.},
  title       = {STRAHL User Manual},
  institution = {Max-Planck-Institut fuer Plasmaphysik},
  number      = {10/30},
  address     = {Garching},
  year        = {2006}
}

@article{Sciortino2021,
  author  = {Sciortino, F. and others},
  journal = {Plasma Physics and Controlled Fusion},
  volume  = {63},
  pages   = {112001},
  year    = {2021}
}

@article{Krasheninnikov2011,
  author  = {Krasheninnikov, S. I. and Smirnov, R. D. and Rudakov, D. L.},
  journal = {Plasma Physics and Controlled Fusion},
  volume  = {53},
  pages   = {083001},
  year    = {2011}
}

@article{Pigarov2005,
  author  = {Pigarov, A. Yu. and Krasheninnikov, S. I. and Soboleva, T. K. and Rognlien, T. D.},
  journal = {Physics of Plasmas},
  volume  = {12},
  pages   = {122508},
  year    = {2005}
}

@article{Obukhov1949,
  author  = {Obukhov, A. M.},
  journal = {Izvestiya Akademii Nauk SSSR, Seriya Geograficheskaya i Geofizicheskaya},
  volume  = {13},
  pages   = {58},
  year    = {1949}
}

@article{Corrsin1951,
  author  = {Corrsin, Stanley},
  journal = {Journal of Applied Physics},
  volume  = {22},
  pages   = {469},
  year    = {1951}
}

@article{Yaglom1949,
  author  = {Yaglom, A. M.},
  journal = {Doklady Akademii Nauk SSSR},
  volume  = {69},
  pages   = {743},
  year    = {1949}
}

@article{ConstantinTiti1994,
  author  = {Constantin, Peter and E, Weinan and Titi, Edriss S.},
  journal = {Communications in Mathematical Physics},
  volume  = {165},
  pages   = {207},
  year    = {1994}
}

@article{DuchonRobert2000,
  author  = {Duchon, Jean and Robert, Raoul},
  journal = {Nonlinearity},
  volume  = {13},
  pages   = {249},
  year    = {2000}
}

@article{Eyink2003,
  author  = {Eyink, Gregory L.},
  journal = {Nonlinearity},
  volume  = {16},
  pages   = {137},
  year    = {2003}
}

@article{WatanabeGotoh2004,
  author  = {Watanabe, T. and Gotoh, T.},
  journal = {New Journal of Physics},
  volume  = {6},
  pages   = {40},
  year    = {2004}
}

@article{Sreenivasan1996,
  author  = {Sreenivasan, K. R.},
  journal = {Physics of Fluids},
  volume  = {8},
  pages   = {189},
  year    = {1996}
}

@article{Aluie2011,
  author  = {Aluie, Hussein},
  journal = {Physical Review Letters},
  volume  = {106},
  pages   = {174502},
  year    = {2011}
}

@article{Aluie2013,
  author  = {Aluie, Hussein},
  journal = {Physica D: Nonlinear Phenomena},
  volume  = {247},
  pages   = {54},
  year    = {2013}
}

@article{GaltierBanerjee2011,
  author  = {Galtier, S. and Banerjee, S.},
  journal = {Physical Review Letters},
  volume  = {107},
  pages   = {134501},
  year    = {2011}
}

@article{BanerjeeGaltier2013,
  author  = {Banerjee, S. and Galtier, S.},
  journal = {Physical Review E},
  volume  = {87},
  pages   = {013019},
  year    = {2013}
}

@article{EyinkDrivas2018,
  author  = {Eyink, Gregory L. and Drivas, Theodore D.},
  journal = {Physical Review X},
  volume  = {8},
  pages   = {011022},
  year    = {2018}
}

@article{DrivasEyink2018,
  author  = {Drivas, Theodore D. and Eyink, Gregory L.},
  journal = {Communications in Mathematical Physics},
  volume  = {359},
  pages   = {733},
  year    = {2018}
}

@article{FeireislGwiazda2017,
  author  = {Feireisl, Eduard and Gwiazda, Piotr and Swierczewska-Gwiazda, Agnieszka and Wiedemann, Emil},
  journal = {Archive for Rational Mechanics and Analysis},
  volume  = {223},
  pages   = {1375},
  year    = {2017}
}

@article{DrivasEyink2017,
  author  = {Drivas, Theodore D. and Eyink, Gregory L.},
  journal = {Journal of Fluid Mechanics},
  volume  = {829},
  pages   = {153},
  year    = {2017}
}

@article{ColomboCrippaSorella2023,
  author  = {Colombo, Marco and Crippa, Gianluca and Sorella, Massimo},
  journal = {Annals of PDE},
  volume  = {9},
  pages   = {21},
  year    = {2023}
}

@article{ArmstrongVicol2025,
  author  = {Armstrong, Scott and Vicol, Vlad},
  journal = {Annals of PDE},
  volume  = {11},
  pages   = {2},
  year    = {2025}
}

@article{BrueDeLellis2023,
  author  = {Bru{\`e}, Elia and De Lellis, Camillo},
  journal = {Communications in Mathematical Physics},
  volume  = {400},
  pages   = {1507},
  year    = {2023}
}

@misc{Companion,
  author       = {Mayboroda, Svitlana and Spergel, David N. and De Lellis, Camillo},
  title        = {The landscape of compressible turbulence},
  year         = {2026},
  note         = {Companion paper}
}

@incollection{Braginskii1965,
  author    = {Braginskii, S. I.},
  title     = {Transport Processes in a Plasma},
  booktitle = {Reviews of Plasma Physics},
  volume    = {1},
  editor    = {Leontovich, M. A.},
  publisher = {Consultants Bureau},
  address   = {New York},
  pages     = {205--311},
  year      = {1965}
}

@article{SpitzerHarm1953,
  author  = {Spitzer, Lyman and H{\"a}rm, Richard},
  journal = {Physical Review},
  volume  = {89},
  pages   = {977},
  year    = {1953}
}

@article{EpperleinHaines1986,
  author  = {Epperlein, E. M. and Haines, M. G.},
  journal = {Physics of Fluids},
  volume  = {29},
  pages   = {1029},
  year    = {1986}
}

@article{dCNChacon2011,
  author  = {del-Castillo-Negrete, D. and Chac{\'o}n, L.},
  journal = {Physical Review Letters},
  volume  = {106},
  pages   = {195004},
  year    = {2011}
}

@article{Gunter2005,
  author  = {G{\"u}nter, S. and Yu, Q. and Kr{\"u}ger, J. and Lackner, K.},
  journal = {Journal of Computational Physics},
  volume  = {209},
  pages   = {354},
  year    = {2005}
}

@article{Holzl2009,
  author  = {H{\"o}lzl, M. and G{\"u}nter, S. and Classen, I. G. J. and Yu, Q. and Delabie, E. and TEXTOR Team},
  journal = {Nuclear Fusion},
  volume  = {49},
  pages   = {115009},
  year    = {2009}
}

@article{Balbus2000,
  author  = {Balbus, Steven A.},
  journal = {Astrophysical Journal},
  volume  = {534},
  pages   = {420},
  year    = {2000}
}

@article{Quataert2008,
  author  = {Quataert, Eliot},
  journal = {Astrophysical Journal},
  volume  = {673},
  pages   = {758},
  year    = {2008}
}

@article{ParrishStone2005,
  author  = {Parrish, Ian J. and Stone, James M.},
  journal = {Astrophysical Journal},
  volume  = {633},
  pages   = {334},
  year    = {2005}
}

@article{SharmaHammett2007,
  author  = {Sharma, Prateek and Hammett, Gregory W.},
  journal = {Journal of Computational Physics},
  volume  = {227},
  pages   = {123},
  year    = {2007}
}

@article{SharmaHammett2011,
  author  = {Sharma, Prateek and Hammett, Gregory W.},
  journal = {Journal of Computational Physics},
  volume  = {230},
  pages   = {4899},
  year    = {2011}
}

@article{ChaconDCNHauck2014,
  author  = {Chac{\'o}n, L. and del-Castillo-Negrete, D. and Hauck, C. D.},
  journal = {Journal of Computational Physics},
  volume  = {272},
  pages   = {719},
  year    = {2014}
}

@article{MFPNAS,
  author  = {Filoche, M. and Mayboroda, S.},
  journal = {Proceedings of the National Academy of Sciences of the United States of America},
  volume  = {109},
  pages   = {14761},
  year    = {2012}
}

@article{DrivasElgindiIyerJeong2022,
  author  = {Drivas, Theodore D. and Elgindi, Tarek M. and Iyer, Gautam and Jeong, In-Jee},
  journal = {Archive for Rational Mechanics and Analysis},
  volume  = {243},
  pages   = {1151},
  year    = {2022}
}

@article{Kritsuk2007,
  author  = {Kritsuk, Alexei G. and Norman, Michael L. and Padoan, Paolo and Wagner, Ralf},
  journal = {Astrophysical Journal},
  volume  = {665},
  pages   = {416},
  year    = {2007}
}

@article{Ni2015,
  author  = {Ni, Q.},
  journal = {Physical Review E},
  volume  = {91},
  pages   = {053020},
  year    = {2015}
}

@article{Ni2016,
  author  = {Ni, Q.},
  journal = {Physical Review E},
  volume  = {93},
  pages   = {043116},
  year    = {2016}
}

@article{dCNChacon2012,
  author  = {del-Castillo-Negrete, D. and Chac{\'o}n, L.},
  journal = {Physics of Plasmas},
  volume  = {19},
  pages   = {056112},
  year    = {2012}
}

@article{RechesterRosenbluth1978,
  author  = {Rechester, A. B. and Rosenbluth, M. N.},
  journal = {Physical Review Letters},
  volume  = {40},
  pages   = {38},
  year    = {1978}
}

\appendix
\section{END MATTER}
We collect the definitions deferred from the main text, the skeleton of the
derivation of \eqref{eq:master}, and the coarse-graining estimates.  The
companion paper \cite{Companion} develops the source-free case in full,
including the solution class, the kernel-free Besov formulation, and the
sharpness example; the source terms and the increment estimates involving $f$
are as derived here.  Every step below is written for the matrix landscape
$\Lm$; setting $\Adif=\bm I$ recovers the isotropic case with constant
diffusivity verbatim.

\section{Landscape averaging and the deferred definitions}
\label{app:defs}

The measure $\dd\nu_{\ell,x}$ of \eqref{eq:S} is a probability measure and
$\tau^\rho_\ell(f,g)=\rho_\ell\,\mathrm{Cov}_{\nu_{\ell,x}}(f,g)$, so
$\tau^\rho_\ell(f,f)\ge0$.  Its landscape counterpart for vector fields is
\begin{equation}
 \lFavre{\bm f}_\ell=\Lm_\ell^{-1}\big(\Lm\bm f\big)_\ell,
 \;\;
 \tau^{\Lm}_\ell(\bm f,\bm g)=\big(\bm f\!\cdot\!\Lm\bm g\big)_\ell
 -\lFavre{\bm f}_\ell\!\cdot\!\Lm_\ell\lFavre{\bm g}_\ell ,
 \label{eq:lfavre}
\end{equation}
and $\tau^{\Lm}_\ell(\bm f,\bm f)\ge0$ now by a \emph{matrix} Cauchy--Schwarz
inequality rather than by positivity of a scalar variance: for any constant
vector $\bm a$,
\begin{equation}
 \big((\bm f-\bm a)\!\cdot\!\Lm(\bm f-\bm a)\big)_\ell
 =\big(\bm f\!\cdot\!\Lm\bm f\big)_\ell
 -2\bm a\!\cdot\!(\Lm\bm f)_\ell
 +\bm a\!\cdot\!\Lm_\ell\bm a\ \ge0 ,
 \label{eq:matrix-CS}
\end{equation}
minimized at $\bm a=\lFavre{\bm f}_\ell$, where its value is exactly
$\tau^{\Lm}_\ell(\bm f,\bm f)$.  Thus $\lFavre{\bm f}_\ell$ is the
$\Lm$-orthogonal projection of $\bm f$ onto constants at scale $\ell$ and
$\tau^{\Lm}_\ell$ is the residual.  This single observation, a completion of the
square, replaces every use of scalar positivity in the isotropic argument and
is the technical heart of the extension.  With a matrix landscape,
$\Am_\ell=\lFavre{(\nabla\theta)}_\ell$ is not a weighted mean of
$\nabla\theta$ component by component but the solution of the linear system
$\Lm_\ell\Am_\ell=(\Lm\nabla\theta)_\ell$, which mixes components wherever
$\bhat$ varies inside the filter; and
\begin{equation}
 \Cm_\ell=(\Lm\nabla\theta)_\ell-\Lm_\ell\nabla h_\ell
 =\Lm_\ell\big(\Am_\ell-\nabla h_\ell\big).
 \label{eq:C}
\end{equation}
Strict positivity of $K_\ell$ together with $\int\rho\,\dd x=1$ gives
$\rho_\ell\ge c_\ell>0$ and hence $\Lm_\ell\succeq a_-c_\ell\bm I\succ0$, with
no lower bound on $\rho$ or on $\Lm$; the constant $c_\ell$ degenerates as
$\ell\downarrow0$ and appears in no estimate below.

\section{Filtered balances and the resolved budget}
\label{app:filtered}

Filtering \eqref{eq:cont}--\eqref{eq:scalar} with $K_\ell$ gives, exactly,
\begin{align}
\partial_t\rho_\ell+\nabla\!\cdot(\rho_\ell\bm V_\ell)&=0,
\label{eq:fcont}\\
\partial_t(\rho_\ell h_\ell)
+\nabla\!\cdot\!\big[\rho_\ell\bm V_\ell h_\ell+\tau^\rho_\ell(\bm v,\theta)\big]
&=\kappa\nabla\!\cdot(\Lm_\ell\Am_\ell)+S_\ell .
\label{eq:filteredscalar}
\end{align}
That \eqref{eq:fcont} is exact, with no subgrid term, is the reason Favre
averaging with $\rho$ is the natural weighting for $\theta$ and $\bm v$; that
$(\Lm\nabla\theta)_\ell$ equals $\Lm_\ell\Am_\ell$ by the definition of
$\Am_\ell$ as the solution of a linear system is the reason the landscape is
the natural weight for the gradient.  Both are exact for an arbitrary
symmetric positive-definite $\Adif(x,t)$.  Multiplying
\eqref{eq:filteredscalar} by $h_\ell$, using \eqref{eq:fcont}, and integrating
over $\mathbb T^d$ gives the resolved budget
\begin{equation}
 \frac{\dd}{\dd t}\frac12\!\int\rho_\ell h_\ell^2\,\dd x
 =-\Pi^\rho_\ell
 -\kappa\!\int\nabla h_\ell\cdot\Lm_\ell\Am_\ell\,\dd x
 +\!\int h_\ell S_\ell\,\dd x ,
 \label{eq:resolvedinstant}
\end{equation}
which is also the starting point of \eqref{eq:stationary-master}: under time
translation invariance the left-hand side has vanishing mean, so
$\mathbb E\Pi^\rho_\ell=\mathbb EI_\ell-\mathbb EJ_\ell$ with
$I_\ell:=\int h_\ell S_\ell\,\dd x$, and
$I_\ell=\int S\theta\,\dd x-\int\tau^\rho_\ell(f,\theta)\,\dd x$ because
convolution preserves spatial integrals.

Because $\tau^{\Lm}_\ell$ is a matrix-weighted covariance in the sense of
\eqref{eq:matrix-CS}, the landscape-weighted gradient variance splits exactly,
\begin{equation}
 \int\nabla\theta\cdot\Lm\nabla\theta
 =\int\Am_\ell\cdot\Lm_\ell\Am_\ell
 +\int\tau^{\Lm}_\ell(\nabla\theta,\nabla\theta),
 \label{eq:gradient-split}
\end{equation}
with both terms on the right nonnegative, while \eqref{eq:C} gives the
pointwise algebraic decomposition
\begin{equation}
 \Am_\ell\cdot\Lm_\ell\Am_\ell
 =\nabla h_\ell\cdot\Lm_\ell\nabla h_\ell
 +2\Cm_\ell\!\cdot\nabla h_\ell
 +\Cm_\ell\cdot\Lm_\ell^{-1}\Cm_\ell .
 \label{eq:A-decomposition}
\end{equation}
The latter follows by substituting $\Am_\ell=\nabla h_\ell+\Lm_\ell^{-1}\Cm_\ell$
and expanding, the cross term being
$2\nabla h_\ell\cdot\Lm_\ell\Lm_\ell^{-1}\Cm_\ell=2\nabla h_\ell\cdot\Cm_\ell$
and the last
$\Cm_\ell\cdot\Lm_\ell^{-1}\Lm_\ell\Lm_\ell^{-1}\Cm_\ell$, both using the
symmetry of $\Lm_\ell$.  Symmetry is used here and in \eqref{eq:matrix-CS} and
nowhere else, which is why an antisymmetric part of the diffusivity is not
covered.  Equations \eqref{eq:gradient-split} and \eqref{eq:A-decomposition}
are the only two facts about the gradient field needed below, and both are
statements about the single positive-definite landscape, indifferent to how it
is built out of $\rho$ and $\Adif$ and in particular to the anisotropy.

\section{Derivation of the balance}
\label{app:master}

Introduce the commutator work and the resolved dissipation,
\begin{equation}
 X:=\kappa\!\int_t^T\!\!\int\Cm_\ell\cdot\nabla h_\ell\,\dd x\,\dd s ,
 \quad
 G:=\kappa\!\int_t^T\!\!\int\Am_\ell\cdot\Lm_\ell\Am_\ell\,\dd x\,\dd s ,
 \label{eq:XG}
\end{equation}
so that $\kappa\int_t^T\!\int\nabla h_\ell\cdot\Lm_\ell\Am_\ell=H+X$ by
\eqref{eq:C}.  Integrating \eqref{eq:resolvedinstant} from $t$ to $T$ and
subtracting the corresponding integral of \eqref{eq:unfilteredenergy} gives,
with the definitions \eqref{eq:B} and \eqref{eq:source-defect},
\begin{equation}
 D=E+H+X ,
 \label{eq:Dfirst}
\end{equation}
for renormalized solutions, and $D\le E+H+X$ whenever
\eqref{eq:unfilteredenergy} holds only as an inequality.  Integrating
\eqref{eq:gradient-split} in time gives
\begin{equation}
 R=D-G\ \ge0 ,
 \label{eq:RDG}
\end{equation}
and integrating \eqref{eq:A-decomposition} gives
\begin{equation}
 G=H+2X+N .
 \label{eq:GHXN}
\end{equation}
Eliminating $G$ between \eqref{eq:RDG} and \eqref{eq:GHXN} gives
$X=\tfrac12(D-R-H-N)$, and substituting into \eqref{eq:Dfirst} yields
$D+R+N=2E+H$, which is \eqref{eq:master}; the inequality
\eqref{eq:masterineq} follows since $R,N\ge0$.  The commutator work $X$ appears
only as an intermediate quantity and is never estimated.  No step in this
section refers to the dimension, to the anisotropy, or to any property of $\Lm$
beyond symmetry and pointwise positive definiteness, and the only place where
the unfiltered fields enter at all is the single use of
\eqref{eq:unfilteredenergy}.

\section{Coarse-graining estimates}
\label{app:estimates}

Since $\int\nabla K_\ell=0$,
\begin{equation}
 \rho_\ell\nabla\Favre f_\ell
 =\int\nabla K_\ell(y)\rho(x-y)\big[f(x-y)-\Favre f_\ell(x)\big]\dd y ,
 \label{eq:favre-gradient-identity}
\end{equation}
so the kernel properties give
\begin{equation}
 |\nabla\Favre f_\ell(x)|
 \le C_K\ell^{-1}\!\int|f(x-y)-\Favre f_\ell(x)|\,\dd\nu_{\ell,x}(y).
 \label{eq:favre-gradient-bound}
\end{equation}
Because $\nu_{\ell,x}$ is a probability measure, no comparison between $\rho$
mollified at neighbouring scales is required and no density-doubling
assumption enters.  Writing
$\tau^\rho_\ell(\bm v,\theta)=\rho_\ell\,\mathrm{Cov}_{\nu_{\ell,x}}(\bm v,\theta)$,
applying \eqref{eq:favre-gradient-bound} to $h_\ell$, and using H\"older with
exponents $(3,3,3)$ in $\nu_{\ell,x}$ and then in $\rho_\ell\,\dd x$ gives
\eqref{eq:fluxlemma}; neither $h_\ell$ nor $\tau^\rho_\ell(\bm v,\theta)$
involves the landscape, so this estimate is literally that of the isotropic
case and is unchanged by the anisotropy.  Replacing H\"older by Jensen, and
using that $\rho_\ell\,\dd x$ is a probability measure so that
$\Srho_{2,\ell}\le\Srho_{3,\ell}$ with no constant, gives the bound on
$B^\rho_\ell$.  For $H$ one uses the pointwise landscape--density comparison
$\Lm_\ell=(\rho\Adif)_\ell\preceq a_+\rho_\ell\bm I$ as quadratic forms, valid
by \eqref{eq:ellipticity} and preserved under convolution with a positive
kernel, to reduce to a $\rho_\ell$-weighted quantity:
\begin{equation}
 H\le\kappa a_+\!\int_t^T\!\!\int\rho_\ell|\nabla h_\ell|^2\,\dd x\,\dd s
 \le C_K^2a_+T\,M_\theta^2\,\kappa\,\ell^{2\beta-2}.
 \label{eq:Hbound}
\end{equation}
Symmetrically $D\ge\kappa a_-\int_t^T\!\int\rho|\nabla\theta|^2$, which is how
\eqref{eq:main} is converted into \eqref{eq:limit}.  Only these two pointwise
matrix comparisons are used; no increment estimate for $\Adif$, for $\bhat$, or
for $\Lm$ is ever needed, which is why the diffusivity may be arbitrarily rough
and arbitrarily anisotropic in direction.  The upper bound in
\eqref{eq:ellipticity} may itself be replaced by the third moment $a_3$ of the
main text, since $|\nabla h_\ell|^2$ is controlled in $L^{3/2}(\rho_\ell\dd x)$
by \eqref{eq:favre-gradient-bound} \cite{Companion}.

Combining \eqref{eq:matrix-CS} in the form
$\int\Am_\ell\cdot\Lm_\ell\Am_\ell\le\int\nabla\theta\cdot\Lm\nabla\theta$ with
Cauchy--Schwarz in the $\Lm_\ell$ inner product and \eqref{eq:Hbound} gives the
bound used for \eqref{eq:constant-flux},
\begin{equation}
 \mathbb E|J_\ell|
 \le(\D^\rho_\kappa)^{1/2}\big(\mathbb E H/(T-t)\big)^{1/2}
 \lesssim(\D^\rho_\kappa)^{1/2}M_\theta\big(a_+\kappa\ell^{2\beta-2}\big)^{1/2},
 \label{eq:J-bound}
\end{equation}
so the molecular correction vanishes whenever
$\sup_\kappa\D^\rho_\kappa<\infty$ and $a_+\kappa\,r_\kappa^{2\beta-2}\to0$,
which is the inertial-range window quoted in the main text.

For the source, \eqref{eq:favre} with $S=\rho f$ gives
$\tau^\rho_\ell(f,\theta)=(\theta S)_\ell-h_\ell S_\ell$, and since convolution
preserves spatial integrals,
\begin{equation}
 \Delta Q_\ell(t,T)=\int_t^T\!\!\int\tau^\rho_\ell(f,\theta)\,\dd x\,\dd s .
 \label{eq:source-defect-cov}
\end{equation}
Weighted Cauchy--Schwarz gives
$|\int\tau^\rho_\ell(f,\theta)\dd x|\le\Srho_{2,\ell}(f)\Srho_{2,\ell}(\theta)$,
which with \eqref{eq:scaling} and the source hypothesis yields the stated bound
on $\Delta Q_\ell$.  This part of the argument uses only $\rho$-weighted
averages and is therefore independent of the landscape.  The source defect is a
consequence of the increment hypotheses, not an independent assumption.  In the
heat-conduction reading \eqref{eq:heat-dictionary} this is the step that fails
at a shock, since $f$ then contains $\nabla\!\cdot\bm v$.

\end{document}